\documentclass[multphys,vecphys]{svmult}

\usepackage{makeidx}         
\usepackage{graphicx}        
\usepackage{multicol}        
\usepackage[bottom]{footmisc}

\makeindex             

\begin{document}

\title*{Recent analytical and numerical techniques applied to
the Einstein equations}
\titlerunning{Recent analytical and numerical techniques} 
\author{Dave Neilsen\inst{1,2}, Luis Lehner\inst{1}, Olivier Sarbach\inst{1,3} and Manuel Tiglio\inst{1,4,5}
}
\institute{Department of Physics \& Astronomy\\
Louisiana State University, Baton Rouge, LA 70803, USA
\and Department of Physics \& Astronomy\\
Brigham Young University, Provo, UT 84602, USA
\and Theoretical Astrophysics 130-33\\
California Institute of Technology, Pasadena, CA 91125, USA
\and Center for Computation and Technology,
Louisiana State University, Baton Rouge, LA 70803, USA \\
\and Center for Radiophysics and Space Research, Cornell University, Ithaca, NY 14853
}

%
%
\maketitle

\begin{abstract}
Combining deeper insight of Einstein's equations with sophisticated
numerical techniques promises the ability to construct accurate
numerical implementations of these equations. We illustrate this in
two examples, the numerical evolution of ``bubble'' and single black
hole spacetimes. The former is chosen to demonstrate how accurate
numerical solutions can answer open questions and even reveal
unexpected phenomena. The latter illustrates some of the difficulties
encountered in three-dimensional black hole simulations, and presents
some possible remedies.
\end{abstract}

\section{Introduction}
\label{Sect:Intro}

Extracting the full physical content from Einstein's equations has
proven to be a difficult task. The complexity of these equations has
allowed researchers only a peek into the rich phenomenology of the
theory by assuming special symmetries and reductions. Computational
methods, however, are opening a new window into the theory. To realize
the full utility of computational solutions in exploring Einstein's
equations, several questions must first be addressed. Namely, a deeper
understanding of the system of equations and its boundary conditions,
the development and use of more refined numerical techniques and an
efficient use of the available computational resources.

In recent years, considerable advances have been made in some of these
issues, allowing for the analysis of complex physical systems which
arguably must be tackled numerically. In the present article we
highlight some recent analytical and numerical techniques and apply
them to two practical applications. The first application is the {\it
numerical evolution of bubble spacetimes in five-dimensional
Kaluza-Klein theory}. We study their dynamical behavior, the validity
of cosmic censorship --in a set-up which a-priori would appear promising
to give rise to violations of the conjecture-- and reveal the existence of critical
phenomena. As a second application, we discuss the {\it numerical
evolution of single black hole spacetimes}. Here we consider some
analytical and numerical difficulties in modeling these systems
accurately. We discuss a method to alleviate some of these problems,
and present tests to demonstrate the promise of this method.

\section{Analytical and Numerical tools}
\label{Sect:AnNumTools}

In the Cauchy formulation of General Relativity, Einstein's field
equations are split into evolution and constraint equations. Numerical
solutions are found by specifying data on an initial spacelike slice,
subject to the constraints, and by integrating the evolution equations
to obtain the future development of the data. Owing to finite computer
resources, one is forced to use finite, and, in practice, rather small
computational domains to discretize the problem. This raises several
important issues.

The fundamental property for any useful numerical solution is that the
solution must convergence to the continuum solution in the limit of
infinite resolution.  A prerequisite for a well-behaved numerical
solution is a well-posed continuum formulation of the initial-boundary
value problem. In certain cases, the well-posed continuum problem can
then be used to construct stable numerical discretizations for which
one can {\it a priori} guarantee convergence. In particular, this can
be achieved for linear, first-order, symmetric hyperbolic systems with
maximally dissipative boundary conditions~\cite{gko,strand,olsson}.
This is briefly discussed in Sec.~\ref{Sect:SBP}, for a detailed
description and an extension to numerical relativity see
Refs.~\cite{paperI,paperII, paperIII,Calabrese:2002ej}.

The application of these ideas in general relativity is, naturally,
more complicated. First, Einstein's equations are nonlinear and so it
is much harder to {\it a priori} prove convergence. However, a
discretization that guarantees stability for the linearized equations
should already be useful for the nonlinear equations, especially for
those systems with smooth solutions as expected for the Einstein
equations when written appropriately. This is because in a small
enough neighborhood of any given spacelike slice, the numerical
solution can be modeled as a small amplitude perturbation of the
continuum solution.

The constraint equations in general relativity bring additional
complications and greatly restrict the freedom in specifying boundary
and initial data. This is illustrated and further discussed in
Section~\ref{Sect:CPBC}. Section~\ref{Sect:DynControl} discusses
issues regarding the stability of the constraint manifold. The
manifold is invariant with respect to the flow defined by the
evolution system in the continuum problem. Numerically, however, small
errors in the solution arising from truncation or roundoff error may
lead to large constraint violations if the constraint manifold is
unstable. Section~\ref{Sect:DynControl} discusses a method for
suppressing such rapid constraint violations.

\subsection{Guidelines for a stable numerical implementation.}
\label{Sect:SBP}

A simple numerical algorithm, or ``recipe,'' can be followed to solve
first order, linear symmetric hyperbolic equations with variable
coefficients and maximally dissipative boundary conditions, for which
stability can be guaranteed. It is based on finite difference approximations with
spatial difference operators that satisfy the {\it summation by parts}
(SBP) property. This property is a discrete analogous of {\it
integration by parts}, which is used in the derivation of energy
estimates, a key ingredient for obtaining a well posed formulation of
the continuum problem. SBP allows to obtain similar energy estimates
for the discrete problem.
 
\subparagraph{Employ spatial difference operators that satisfy SBP on the 
computational domain.}

For the sake of simplicity, consider a set of linear, first order
symmetric hyperbolic equations in the one-dimensional domain $x \in
(a,b)$ which is discretized with points $x_j= a + j\Delta x$,
$j=0\ldots N$, where $\Delta x=(b-a)/N$. Now let us introduce the
discrete scalar product,
\begin{equation} 
(u,v) := \Delta x \sum_{i,j=0}^N \sigma_{ij} u_i v_j \, , 
\end{equation} 
for some positive definite matrix with elements $\sigma_{ij}$ which in
the continuum limit $\Delta x \to 0$ approaches the $L_2$ norm
$\langle u,v\rangle := \int_a^b uv \,dx $. At the continuum level, the
derivative operator $d/dx$ and scalar product satisfy integration by
parts, i.e.  $\langle du/dx,v\rangle + \langle u,dv/dx\rangle = uv|_a^b $,
which in the discrete case is translated into a finite difference
operator $D$ which satisfies $(Du,v) + (u,Dv) = uv|_a^b$ and
approaches $d/dx$ in the continuum limit. The simplest difference
operator and scalar product satisfying SBP are
\begin{equation} 
\begin{array}{lll} 
Du = (u_{i+1}-u_{i})/\Delta x\,,\quad & \sigma_{00}=\frac{1}{2}\quad & \mbox{ for } i=0\\ 
Du = (u_{i+1}- u_{i-1})/(2\Delta x)\,,\quad & \sigma_{ii}=1          \quad & \mbox{ for } i=1\ldots N-1\\ 
Du = (u_{i}-u_{i-1})/\Delta x\,,\quad & \sigma_{NN}=\frac{1}{2}\quad & \mbox{ for } i=N 
\end{array} \label{simpled}
\end{equation} 
where the scalar product is diagonal: $\sigma_{ij}=0$ for $i\ne j$. 
Higher order operators satisfying SBP have been constructed by
Strand~\cite{strand}. Additionally, when dealing with non-trivial
domains containing inner boundaries, additional complexities must be
addressed to attain SBP, see Ref.~\cite{paperI}. The finite operator
$D$ is then used for the discretization of the spatial derivatives in
the evolution equations, thus obtaining a semi-discrete system.

\subparagraph{Impose boundary conditions via orthogonal projections~\cite{olsson}.}
This ensures the consistent treatment of the boundaries, guaranteeing
the correct handling of modes propagating towards, from and tangential
to the boundaries. An energy estimate can be obtained for the
semi-discrete system.

\subparagraph{Implement an appropriate time integration algorithm.}
The resulting semi-discrete system constitutes a large system of ODE's which
can be numerically solved by using a time integrator that satisfies
an energy estimate~\cite{kreiss-wu,tadmor}.

\subparagraph{Consider adding explicit dissipation}
 
It is well known that finite difference approximations do not adequately
represent the highest frequency modes on a given grid,
corresponding to the shortest possible wavelengths that can be
represented on the grid.  If the smallest spacing between
points is $\Delta$, the shortest wavelength is $\lambda_{min} =\Delta$
with the corresponding frequency $k_{max} = 2\pi/
\lambda_{min}$.  These modes can, and often do, travel in the wrong
direction. For this reason, it is sometimes useful to add explicit
numerical dissipation to rid the simulation of these modes in a way
that is consistent with the continuum equation at hand. As
finer grids are used, the effect of this dissipation becomes smaller
and acts only on increasingly higher frequencies. The dissipation operators
are constructed  such that discrete energy estimates, obtained using SBP,
are not spoiled.  Explicit expressions for such dissipation operators are 
presented in Ref.~\cite{paperI}.

To summarize, beginning with a well-posed
initial-boundary value problem, we mimic the derivation of continuum 
energy estimates for the discrete problem using (1) spatial
derivative operators satisfying summation by parts, (2) orthogonal
projections to represent boundary conditions and (3) choosing an
appropriate time integrator.

\subsection{Constraint-preserving boundary conditions}
\label{Sect:CPBC}

As discussed above, a numerical implementation of any system of
partial differential equations necessarily involves boundaries. Unless
periodic boundary conditions can be imposed, as is often the case for
the evolution on compact domains without boundaries, one deals with an
initial-boundary value problem, and thus has to face the question of
how to specify boundary conditions. In theories that give rise to
constraints, like general relativity, such conditions must be chosen
carefully to ensure that the constraints propagate.

As a very simple illustration, consider the 1d wave equation
$u_{,tt}=u_{,xx}$ on the half line $x > 0$. Let us reduce it to first
order form by introducing the variables $f \equiv u_{,x}$ and $g
\equiv u_{,t} - b u_{,x}$, with $b$ a negative constant:
\begin{eqnarray}
u_{,t} &=& b\, u_{,x} + g,
\label{Eq:ut2}\\
g_{,t} &=& - b\, g_{,x} + (1-b^2) f_{,x}\; ,
\label{Eq:gt2}\\
f_{,t} &=& g_{,x} + b\, f_{,x}\; .
\label{Eq:ft2}
\end{eqnarray}
At the boundary $x=0$, the system has two ingoing fields, given by $u$
and $v_{in} \equiv g + b\, f - f$, and one outgoing field. However,
the ingoing fields cannot be given independently, as we see next. The
constraint $C\equiv f-u_{,x} = 0$ propagates as $C_{,t} = b\, C_{,x}$
and so $C$ is an ingoing field with respect to $x=0$. Therefore, we
have to impose the boundary condition $C=0$ which implies the
condition $u_{,x} = f$ on the main variables. We can replace this with
a condition that is intrinsic to the boundary by using the evolution
equation (\ref{Eq:ut2}) in order to eliminate the $x$-derivative and
obtain
\begin{equation}
u_{,t} = b\, f + g.
\end{equation}
This equation provides an evolution equation for determining $u$ at
the boundary, which guarantees that the constraint $C=0$ is preserved
throughout evolution. It can be complemented by the Sommerfeld
condition $v_{in} = 0$.

This simple example gives just a glimpse of the different issues
involved in prescribing constraint-preserving boundary conditions.
The case of Einsteins's field equations is more complicated; we refer
the interested reader to Refs.~\cite{CPBC-Stu, FN, CPBC-IR, CPBC-SSW,
CPBC-BB, CPBC-CLT, CPBC-SW, CPRST, CS, CPBC-Frittelli, GMG, LS-FatMax,
RS-FatMax}. A major difficulty is the fact that, in general,
constraint-preserving boundary conditions do not have the form of
maximal dissipative boundary conditions, and for this reason it has
proven to be difficult to find well posed initial-boundary value
formulations of Einstein's equations that preserve the constraints.

\subsection{Dealing with ``too many'' formulations. 
Parameters via constraint monitoring}
\label{Sect:DynControl}

Formulations of the Einstein equations are often cast in symmetric
hyperbolic form by adding constraints to the evolution equations
multiplied by parameters or spacetime functions. The symmetric
hyperbolicity condition partially restricts these parameters, however,
considerable freedom in the formulation exists in choosing these free
parameters (see, for instance,~\cite{kst}). Analytically, when data are on the constraint
surface, all allowed values for these parameters are equally
valid. Off of the constraint surface, however, different values of
these parameters may be regarded as representing ``different''
theories. It is no surprise then that numerical simulations are
sensitive to the values chosen for these parameters, as numerical data
rarely are on the constraint surface.  Unfortunately, the parameters
in current simulations are proving to be extremely sensitive. Relatively mall
variations in these parameters (within the allowed range for a
symmetric hyperbolic formulation) produce run times in simulations
that vary over several orders of magnitude, as measured by an
asymptotic observer.

Furthermore, the parameters are not unique. Values convenient for one
physical problem might be inappropriate in another.  Recently, a
method to dynamically choose these parameters---promoted to functions
of time---was introduced that naturally adapts to the physical problem
under study~\cite{dyn}. Basically, one exploits the freedom in
choosing these functions to control the growth rate of an energy norm
for constraint violations.  Since this norm is exactly zero
analytically, this provides a guide to choosing the parameters that
will drive the solution to one that satisfies the constraints.
This method provides a {\it practical} solution to this problem of
choosing parameters, although it may not be the most elegant solution.
Ideally, one would like to understand how the growth rate of the solution
depends on the these parameter values in order to choose them appropriately. 
This would require sharp growth
estimates, however, which are still unavailable. While further
understanding is gained in this front, this practical remedy can be of
much help in present simulations. We summarize here the essential
ideas of this method.

Consider a system of hyperbolic equations with constraint terms,
$C_c$, written schematically as
\begin{equation}
\dot{u}_a = \sum_bA^b(u,t,\vec{x})\partial_b u_a
    + B_a(u,t,\vec{x}) + \sum_c\mu_{ac} C_c(u,\partial_ju)
\label{linearc}\;,
\end{equation}
where $u_a$, $B_a$ and $C_c$ are vector valued functions, and
$\mu_{ac}$ is a matrix (generally not square) that is a function of
the spacetime ($C_c$ represents a vector function of general
constraint variables).  The indices $\{a,b,c\}$ range over each
element of the vector or matrix functions, while the indices
$\{i,j,k\}$ label points on a discrete grid.  We define an {\it
energy} or {\it norm} of the discrete constraint variables as
\begin{equation}
{\cal N}(t) = \frac{1}{2n_xn_yn_z} \sum_{c}\sum_{ijk} C_c(t) ^2 ,
\label{constraintnorm}
\end{equation}
where $n_x$, $n_y$, $n_z$ are the number of points in each direction.  The
grid indices $\{i,j,k\}$ are suppressed to simplify the notation.
The time derivative of the norm can be calculated using Eq.~(\ref{linearc})
\begin{equation}
\dot{\cal N} = {\cal I}^{hom}
 + \mbox{Tr} (\mu   {\cal I}^{\mu }), 
\label{split}
\end{equation}
and therefore can be known in closed form provided the
matrix valued sums
\begin{eqnarray}
 {\cal I}^{hom} &=&
\sum_{ijk} \sum_{a,b} \frac{C_a}{n_xn_yn_z}\left[\frac{\partial C_a}{\partial u_b}+
\sum_k\frac{\partial C_a}{\partial D_k u_a}D_k \right] \times
 \nonumber \\
&&  \left[\sum_c(A^cD_cu_b) +B_b\right]  \label{split1}  \\
{\cal I}^{\mu}_{bc} &=&  \sum_{ijk} \sum_{a}
    \frac{C_a}{n_xn_yn_z} \times  \nonumber \\
&& \left[\frac{\partial C_a} {\partial u_b}+
 \sum_k\frac{\partial C_a}{\partial D_k u_b}D_k \right]C_c \label{split2}
\end{eqnarray}
are computed during evolution. Here $D_i$ is the discrete derivative
approximation to $\partial_i$. We then use the dependence of the
energy growth on the free constraint-functions to achieve some desired
behavior for the constraints, i.e., solving Eq.~(\ref{split}) for
$\mu_{ac}$.  For example, if we choose\footnote{There is a slight
abuse of notation here, in the sense that $a$ does not denote an
index, as before. Similarly, the subscript in $n_a$ indicates that the
quantity is related to $a$ through Eq.~(\ref{a}).}
\begin{equation}
\dot{\cal N} = -a {\cal N}, \qquad a>0,   \label{edot}
\end{equation}
any violation of the constraints will decay exponentially
\begin{equation}
{\cal N}(t+\triangle t) = {\cal N}(t)e^{-a\triangle t} \label{decay}\;.
\end{equation}
As discussed in Ref.~\cite{dyn}, one good option among many others seems to
be choosing a tolerance value, $T$, for the norm of the constraints that
is close to the initial discrete value, and solving for $\mu_{ac}$
such that the constraints decay to this tolerance value after a given
relaxation time.  This can be done by adopting an  $a$ such that after some time
$\tau \equiv n_a \triangle t$ the constraints have the value $T$. Replacing ${\cal
N}(t+\triangle t)$ by $T$ in equation (\ref{decay}) and solving for
$a$ gives
\begin{equation}
a(t) = -\frac{1}{\tau}\ln{\left(\frac{T}{{\cal N}(t)}\right)}
\; . \label{a}
\end{equation}
If one then solves
\begin{equation}
\dot{\cal N} = -a {\cal N} = {\cal I}^{hom}
   + \mbox{trace}(\mu \times {\cal I}^{\mu }) \label{eq_for_mu}
\end{equation}
for $\mu$, with $a$ given by Eq.~(\ref{a}), the value of the norm ${\cal
N}(t+\tau)$ should be $T$, independent of its initial value.
Therefore, Eq.~(\ref{eq_for_mu}) serves as a guide to formulate a practical method
to choose free parameters in the equations with which the numerical solution
behaves well with respect to the satisfaction of the constraints. Naturally,
if one deals, as it is often the case, with more than one free parameter, Eq.~(\ref{eq_for_mu})
must be augmented with other conditions to yield a unique solution. This extra
freedom is actually very useful in preventing large time-variations in the parameters
that are sometimes needed in order to keep the constraints under
control. These large variations do not represent a fundamental problem but a
practical one, due to the small time stepping that they require in order to keep
errors due to time integration reasonably small. 
One way to prevent this is by using this extra freedom to pick up the point in
parameter space that not only gives the desired constraint growth, but also
minimizes the change of parameters between two consecutive timesteps. 

Rather than including the full details on the particular way we have
implemented the method, we describe here a simple example to 
illustrate its application. Assume, for instance,
that within a particular formulation only two free functions, $\{\kappa,\omega\}$, 
are employed, Eq.~(\ref{eq_for_mu}) formally evaluates to
\begin{equation}
\dot{\cal N} = -a {\cal N} = {\cal I}^{hom}
   + \kappa {\cal I}^{\kappa} + \omega {\cal I}^{\omega} \label{eq_for_mu_ex}. 
\end{equation}
Now, we exploit the freedom in the free functions to adjust the 
rate of change of
the energy ${\cal N}$ if the values of $\{{\cal I}^{hom},{\cal I}^{\kappa},{\cal I}^{\omega}\}$ are
known. In practice, these  are easily obtained during evolution. 
Once these are known, Eq.~(\ref{eq_for_mu_ex}),
coupled to the requirement that $\{\kappa,\omega\}$ vary as little as possible
from one evaluation
to another, 
results in a straightforward strategy to evaluate preferred values of the free
parameters. This is done at a single
resolution ``test'' run and, through interpolation in time, continuum, 
{\it a priori}
defined parameters which keep the constraints under control for the given
problem are obtained. Depending on the formulation of the equations,
the free parameters might have to satisfy some conditions in order for
symmetric hyperbolicity to hold, which can restrict the range of
values these parameters can take. Nevertheless, even within a restricted 
window, the technique allows
one to adopt the most convenient values these parameters should have for 
the problem at hand.

\section{Applications}
\label{Sect:Appl}

We now present applications of the techniques previously
discussed. The goal is to illustrate how well-resolved simulations can
indeed serve as a powerful tool to understand particular problems. To
this end we have chosen a problem found in higher dimensional general
relativity. A second application is that of the simulation of single
black hole spacetimes, where the issue of the {\it a priori} lack of a
preferred formulation  is illustrated.

\subsection{Bubble spacetimes}

As a first application we concentrate on the study of {\it bubble
spacetimes} and elucidate the dynamical behavior of configurations
with both positive and negative masses and their possible connection
to naked singularities. Bubble spacetimes have been studied
extensively within five-dimensional Kaluza-Klein theory. These are
five-dimensional spacetimes in which the circumference of the
``extra'' dimensions shrinks to zero on some compact surface referred
to as the ``bubble''.  These bubbles were initially studied by their
relevance in the quantum instability of flat spacetime \cite{witten},
as bubbles can be obtained via semi-classical tunneling from it. They
were later extended to include data corresponding to negative energy
configurations (at a moment of time
symmetry)~\cite{brillpfister,brillhorowitz}. As mentioned, among the
reasons for considering negative energy solutions is that naked
singularities are associated with them. Therefore, these solutions
are attractive tests of the cosmic
censorship conjecture. Additionally, bubble spacetimes can also be
obtained by double-Wick rotation of black strings, whose stability
properties (or lack thereof) have been the subject of intense scrutiny
in recent years.  These features make bubble spacetimes both interesting an
relevant for gravity beyond four-dimensions, and thus attention has been
devoted to fully understand their behavior. As we will see, even when
the ``analytical'' study of the problem is greatly simplified by
symmetry assumptions, many lingering questions remain and numerical
simulations provided a viable way to shed light into
them. Furthermore, these simulations were also key to `digging out' a
few unexpected features of the solution.

In order to obtain a complete description of the dynamical behavior of
these spacetimes, a numerical code, implementing Einstein equations in
5D settings, and capable of handling the possibly strong curvature
associated need be constructed. Fortunately, the assumption of a ${\bf
SO(3) \times U(1)}$ symmetry simplifies the treatment of the problem,
which can be reduced to a $1+1$ manifold. This, in turn, renders the
problem quite tractable by the currently available computational
resources, though as we will see, considerable care must be placed at
both analytical and numerical levels for an accurate treatment of the problem.

\subsubsection{Initial data}

We consider a generalization of the time symmetric family of initial
data presented in~\cite{brillhorowitz}. We start with a spacetime
endowed with the metric
\begin{equation}
ds^2 = -dt^2 + U(r) dz^2 + \frac{dr^2}{U(r)} + r^2 d\Omega^2 ,
\label{Eq:BubbleMetric}
\end{equation}
where $d\Omega^2 = d\vartheta^2 + \sin^2\vartheta d\varphi^2$ is the
standard metric on the unit two-sphere $S^2$ and $U(r)$ is a smooth
function that has a regular root at some $r=r_+ > 0$, is everywhere
positive for $r > r_+$ and converges to one as $r\rightarrow\infty$.
The coordinate $z$ parameterizes the extra dimension $S^1$ which has
the period $P = 4\pi/U'(r_+)$. The resulting spacetime $\{ t,z,r\geq
r_+,\vartheta,\varphi \}$ constitutes a regular manifold with the
topology $R \times R^2\times S^2$.  The bubble is located where
the circumference of the extra dimension shrinks to zero, that is, at
$r=r_+$.

Additionally, we consider the presence of an electromagnetic field of
the form
\begin{equation}
\frac{1}{2}\, F_{\mu\nu}\, dx^\mu \wedge dx^\nu = d\gamma(r) \wedge dz,
\label{Eq:BubbleEMField}
\end {equation}
where $\gamma(r)$ is a smooth function of $r$ that converges to zero
as $r\rightarrow\infty$. The symmetries of the problem would also
allow for a non-trivial electric component of the field. However, it
is not difficult to show that Maxwell's equations imply that such a
field necessarily diverges at the location of the bubble. For this
reason, in the following, we only consider the case of vanishing
electric field.

In this article, we consider initial data with
\begin{equation}
\gamma(r) = k (r_+^{-n} - r^{-n}),
\label{Eq:IDgamma}
\end{equation}
where $k$ is an arbitrary constant and $n$ an integer greater than
one. This field generalizes the ansatz considered
in~\cite{brillhorowitz}, where only the case $n=2$ was discussed, and
allows for different interesting initial configurations. In the
time-symmetric case, initial data satisfying the Hamiltonian
constraint obeys
\begin{equation}
U(r) = 1 - \frac{m}{r} + \frac{b}{r^2} - \frac{\tilde k^2}{r^{2n}}\, ,
\label{Eq:IDU}
\end{equation} 
with $\tilde{k} \equiv k n/\sqrt{(n-1)(2n-1)}$ and free integration
constants $m$ and $b$. Here, the parameter $m$ is related to the ADM
mass via $M_{ADM} = m/4$. The fact that the bubble be located at
$r=r_+$ requires that $0 = U(r_+) = 1 - \bar{m} + \bar{b} -
\bar{k}^2$, where $\bar{m} \equiv m/r_+$, $\bar{b} \equiv b/r_+^2$,
$\bar{k} \equiv \tilde{k}/r_+^n\,$. We also require
\begin{equation}
0 < r_+ U'(r_+) = 2 - \bar{m} + 2(n-1)\bar{k}^2
\label{Eq:RegHor}
\end{equation}
and avoid the conical singularity at $r=r_+$ by fixing the period of
$z$ to $P = 4\pi/U'(r_+)$. It can be shown that the initial acceleration
of the bubble area $A$ with respect to proper time is given by
\begin{equation}
\ddot{A} = 8\pi\left[ 1 - \bar{m} - \frac{4\bar{k}^2}{3}(n-1)(n-2) \right].
\label{Eq:ddotA}
\end{equation}
For $n=2$, as discussed in Ref.~\cite{corleyjacobson}, this implies
that negative mass bubbles start out expanding (the initial velocity
of the area is zero since we only consider time-symmetric initial
data), while for large enough positive mass the bubble starts out
collapsing. In the vacuum case, our numerical simulations suggest that
initially collapsing bubbles undergo complete collapse and form a
black string. In the non-vacuum case however, the strength of the
electromagnetic field can modify this behavior completely. We will see
that for small enough $k$ the bubble continues to collapse whereas
when $k$ is large the bubble area bounces back and
expands. Interesting behavior is obtained at the critical value for
$k$ which divides the phase space between collapsing and expanding
solutions.

For $n > 2$ it is possible to obtain initial configurations with
negative mass and negative initial acceleration~\cite{SL-Rapid}.  This
can potentially give rise to a collapsing bubble of negative energy,
and thus to a naked singularity. However, our numerical results~\cite{SL-Rapid}
suggest that cosmic censorship is valid: The bubble
bounces back and starts out expanding.

\subsubsection{Equations}

In order to study the time evolution of the initial data sets given on
a $t=const$ slices of the metric (\ref{Eq:BubbleMetric}) and the
electromagnetic field (\ref{Eq:BubbleEMField}), it is convenient to
introduce a new radial coordinate $R = R(r)$ which facilitates the
specification of regularity conditions at the bubble location. This
new coordinate is defined by
\begin{equation}
R(r) = \sqrt{r^2 - r_+^2}\; ,
\qquad   r > r_+\; .
\end{equation}
The metric (\ref{Eq:BubbleMetric}) now reads
\begin{equation}
ds^2 = -\alpha^2 dt^2 + e^{2a} dR^2 + \frac{R^2}{r_+^2 + R^2}\, e^{2b} dz^2 
 + (r_+^2 + R^2) e^{2c} d\Omega^2  ,
\label{Eq:MetricGeneral}
\end{equation}
with $\alpha=1$, $e^{-2a} = e^{2b} = (r_+^2 + R^2)U(R)/R^2$, $c=0$.
Since $U(R) = const\cdot(R/r_+)^2 + O(R^4)$ near $R=0$, and $U(R)$
converges to one in the asymptotic region, $a$ and $b$ are regular
functions. An explicit example is the initial data corresponding to
the zero mass Witten bubble~\cite{witten} where $U = 1 - (r_+/r)^2$
and thus $a=b=0$. When studying the time evolution of the initial data
sets discussed above, we consider the metric (\ref{Eq:MetricGeneral})
where $\alpha$, $a$, $b$ and $c$ are functions of $t$ and $R$. As we
will see, the coordinate $R$ is well suited for imposing regularity
conditions at the bubble location since $(R,z)$ represent polar
coordinates near the bubble, $R=0$ being the center, and $z$ assuming
the role of the angular coordinate. In order to avoid a conical
singularity, $z$ must have the period $2\pi r_+ e^{a-b}$. For this to
be constant we need to impose the boundary condition
$a(t,0)-b(t,0)=const$ at $R=0$.

Similarly, the electromagnetic field (\ref{Eq:BubbleEMField}) is
written in the form
\begin{equation}
\frac{1}{2}\, F_{\mu\nu}\, dx^\mu \wedge dx^\nu 
 = \frac{R}{\sqrt{r_+^2 + R^2}} e^{b}\left( \pi_\gamma dt + d_\gamma dR \right) \wedge dz,
\label{Eq:FieldPrm}
\end{equation}
where the functions $\pi_\gamma$ and $d_\gamma$ depend on $t$ and $R$
and satisfy $\pi_\gamma=0$ and $d_\gamma = e^{-b}\partial_r\gamma$ at
the initial time.

We choose the following gauge condition for the lapse
\begin{equation}
\log(\alpha) = a + \lambda(b + 2c),
\label{Eq:GaugeCond}
\end{equation}
with a parameter $\lambda$ which, in our simulations, is either zero
or one. For $\lambda=1$ the resulting gauge condition is strongly
related to the densitized lapse condition often encountered in
hyperbolic formulations of Einstein's equations: Indeed, the square
root of the determinant of the four metric belonging to
Eq.~(\ref{Eq:MetricGeneral}) is given by $\sqrt{g^{(4)}} = e^{a + b +
2c} R \sqrt{r_+^2 + R^2}\,\sin\vartheta$, so Eq.~(\ref{Eq:GaugeCond})
sets $\alpha$ equal to the square root of the determinant of the four
metric but divides the result by the factor $R \sqrt{r_+^2 +
R^2}\,\sin\vartheta$ which is singular at the bubble, at the poles
$\vartheta=0,\pi/2$ and in the asymptotic region. For $\lambda=0$, the
condition (\ref{Eq:GaugeCond}) implies that the two-metric $-\alpha^2
dt^2 + e^{2a} dR^2$ is in the conformal flat gauge. As we will see,
the principal part of the evolution equations is governed by the
d'Alembertian with respect to this metric. Since the two-dimensional
d'Alembertian operator is conformally covariant, the resulting
equations are semi-linear in that case. In particular, this implies
that the characteristic speeds do not depend on the solution that is
being evolved.

The field equations resulting from the five-dimensional
Einstein-Maxwell equations split into a set of evolution equations and
a set of constraints. The evolution equations can be written as
\begin{eqnarray}
\ddot{A} &=& e^{-\lambda F} \left[ (A' + 2G') e^{\lambda(2 B + F)} \right]'
 - 3(\lambda-1)(C'+G')^2 e^{2\lambda B} - (\lambda+1) V
\nonumber\\
 &+& 2\lambda \dot{A}\dot{B} -\lambda(\lambda+1)\dot{B}^2 - 3(\lambda+1)\dot{C}^2
  +  G\left[ (1-\lambda) \pi_\gamma^2 - (1+\lambda) e^{2\lambda B} d_\gamma^2 \right],
\label{Eq:AMax}\\
\ddot{B} &=& e^{(\lambda-1)B - (\lambda+2) F} \left[ B' e^{(\lambda+1)B + (\lambda+2) F} \right]'
 + \frac{3r_+^2 + 2R^2}{(r_+^2 + R^2)^2}\, e^{2\lambda B}
\nonumber\\
 &+& (\lambda-1)\dot{B}^2 - 2 V,
\label{Eq:BMax}\\
\ddot{C} &=& e^{(\lambda-1)B - F} \left[ (C'+G') e^{(\lambda+1)B + F} \right]' - V
 + (\lambda-1)\dot{B}\dot{C},
\nonumber\\
 &+& \frac{2G}{3}\left[ \pi_\gamma^2 - e^{2\lambda B} d_\gamma^2 \right],
\label{Eq:CMax}\\
\dot{\pi}_\gamma &=& e^{\lambda B - 2(C+G)}\left[ d_\gamma  e^{\lambda B + 2(C+G)} \right]'
  + (\lambda\dot{B} - 2\dot{C}) \pi_\gamma\; ,\\
\dot{d}_\gamma &=& \frac{\sqrt{r_+^2+R^2}}{R} e^{-(B - 2C)}\left[ \pi_\gamma \frac{R}{\sqrt{r_+^2 + R^2}} e^{B-2C} \right]'
  - (\dot{B} - 2\dot{C}) d_\gamma\; ,
\end{eqnarray}
where we have set $A = a + \lambda b + 2(\lambda+1)c$, $B = b + 2c$,
$C = c$ and $G = \log(r_+^2 + R^2)/2$, $F = \log(R) + G$ and $V =
e^{2(A-3C)}/(r_+^2 + R^2)$. Here, and in the following, a dot and a
prime denote differentiation with respect to $t$ and $R$,
respectively.  The evolution equations constitute a hyperbolic system
on the domain $R > 0$.

The constraints are the Hamiltonian and the $R$ component of the
momentum constraint, given by ${\cal C}=0$, ${\cal C}_R = 0$, where
\begin{eqnarray}
{\cal C} &=& e^{(\lambda-1)B - (\lambda+2) F} \left[ e^{(\lambda+1)B + (\lambda+2) F} B' \right]'\nonumber\\
 &+& \left[ \frac{3r_+^2 + 2R^2}{(r_+^2 + R^2)^2} - (B' + F')(A' + 2G') + 3(C'+G')^2 \right] e^{2\lambda B}
\nonumber\\
 &-& V - (\dot{A} - \lambda\dot{B})\dot{B} + 3\dot{C}^2 + G\left[ \pi_\gamma^2 + e^{2\lambda B} d_\gamma^2 \right],
\label{Eq:DefC}\\
{\cal C}_R &=& e^{A - 2C}\left[ e^{-(A- 2C)} \dot{B} \right]' 
 - (B' + F')\left[ \dot{A} - (\lambda+1)\dot{B} \right] + 2(C'+G')(3\dot{C} - \dot{B})
\nonumber\\
 &+& 2G\pi_\gamma d_\gamma\, .
\label{Eq:DefCR}
\end{eqnarray}

\paragraph{Regularity conditions}

The evolution equations contain terms proportional to $e^{-F}$ which
diverge like $1/R$ near $R=0$, and therefore, regularity conditions
have to be imposed at $R=0$. This is achieved by demanding the
boundary conditions
\begin{equation}
A'=B'=C'=\pi_\gamma=0\quad \hbox{at $R=0$}.
\label{Eq:RegCond}
\end{equation}
Assuming that the fields are smooth enough near $R=0$, it then follows
that the right-hand side of the evolution equations is bounded for $R
\rightarrow 0$. Next, as discussed above, the avoidance of a conical
singularity at $R=0$ requires that $A - (\lambda+1)B = a - b$ must be
constant at $R=0$. We show that this condition is a consequence of the
evolution and constraint equations, and of the regularity conditions
(\ref{Eq:RegCond}). Using the evolution equations in the limit
$R\rightarrow 0$ and taking into account the conditions
(\ref{Eq:RegCond}), we find
\begin{equation}
\left. \partial_t \left\{ e^{(1-\lambda)B} \left[ \dot{A} - (\lambda+1)\dot{B} \right] \right\} \right|_{R=0} 
 = \left. -(\lambda+1) e^{(1-\lambda)B} {\cal C} \right|_{R=0} .
\end{equation}
This means that if the Hamiltonian constraint is satisfied at $R=0$
(or in the case that $\lambda=-1$ even if the constraints are
violated), the condition $\left. A - (\lambda+1)B \right|_{R=0} =
const$ will hold provided that the initial data satisfies
$\left. \dot{A} - (\lambda+1)\dot{B} \right|_{R=0} = 0$. Next, we
analyze the propagation of the constraint variables ${\cal C}$ and
${\cal C}_R$ and show that the regularity conditions
(\ref{Eq:RegCond}) and the evolution equations imply that the
constraints are satisfied at each time provided they are satisfied
initially.

\paragraph{Propagation of the constraints}

First, we notice that the vanishing of the momentum constraint
requires that $\left. \dot{A} - (\lambda+1)\dot{B} \right|_{R=0} = 0$
because of the factor $F'$ which diverges like $1/R$ near $R=0$ in the
definition of ${\cal C}_R$. This is precisely the condition
$a(t,0)-b(t,0)=const$ discussed above. However, for this condition to
hold, we first have to show that the momentum constraint actually
vanishes. In order to see this, we regularize the constraint variables
and define $\tilde{\cal C} = e^F {\cal C}$, $\tilde{\cal C}_R = e^F
{\cal C}_R$. Now the regularity conditions (\ref{Eq:RegCond}) imply
that $\tilde{\cal C}_R$ is regular and that $\tilde{\cal C}$ vanishes
at $R=0$. As a consequence of the evolution equations and Bianchi's
identities, the constraint variables obey the following evolution
system
\begin{eqnarray}
\partial_t\tilde{\cal C} &=& 
 e^{(\lambda-1)B}\partial_R\left[ e^{(\lambda+1)B}\tilde{\cal C}_R \right] + (3\lambda-1)\dot{B}\tilde{\cal C}, 
\label{Eq:EvConstr1}\\
\partial_t\tilde{\cal C}_R &=& 
 e^{-(\lambda+1)B - \lambda F}\partial_R\left[ e^{(\lambda+1)B + \lambda F}\tilde{\cal C} \right] 
 + (\lambda-1)\dot{B}\tilde{\cal C}_R
\label{Eq:EvConstr2}
\end{eqnarray}
which is regular at $R=0$. Defining the energy norm
\begin{equation}
{\cal E}(t) = \frac{1}{2} \int_0^\infty 
\left( e^{2B + \lambda F}\tilde{{\cal C}}^2 + e^{2(\lambda+1)B + \lambda F}\tilde{{\cal C}}_R^2 \right) dR,
\end{equation}
taking a time derivative and using the equations
(\ref{Eq:EvConstr1}),(\ref{Eq:EvConstr2}) we obtain
\begin{equation}
\frac{d}{dt}\, {\cal E} 
 = \left. e^{2(\lambda+1)B + \lambda F}\tilde{{\cal C}}\tilde{{\cal C}}_R\right|_0^\infty
  + \lambda\int_0^\infty \dot{B} \left( 3e^{2B + \lambda F}\tilde{{\cal C}}^2 
  + 2e^{2(\lambda+1)B + \lambda F}\tilde{{\cal C}}_R^2 \right) dR.
\label{Eq:EnergyEstConstr}
\end{equation}
The boundary term vanishes because of the regularity conditions at
$R=0$ and under the assumptions that all fields fall off sufficiently
fast as $R\rightarrow\infty$. If $\dot{B}$ is smooth and bounded, we
can estimate the integral on the right-hand side by a constant $C$
times ${\cal E}$, and it follows that ${\cal E}(t) \leq e^{C t}{\cal
E}(0)$. This shows that if the constraints are satisfied initially,
they are also satisfied for all $t>0$ for which a smooth solution
exists. In the gauge where $\lambda=0$ we even obtain the result that
the norm of the constraints cannot grow in time.

To summarize, the boundary conditions (\ref{Eq:RegCond}) imply that
the constraints ${\cal C}=0$, ${\cal C}_R=0$ and $\left. \dot{A} -
(\lambda+1)\dot{B} \right|_{R=0} = 0$ are preserved throughout
evolution.

\paragraph{Outer boundary conditions}

For numerical computations, our domain extends from $R=0$ to
$R=R_{max}$ for some $R_{max} > 0$. Now we have to replace the
estimate (\ref{Eq:EnergyEstConstr}) by the estimate
\begin{equation}
\frac{d}{dt}\, {\cal E} 
 = \left. e^{2(\lambda+1)B + \lambda F + A_0}\tilde{{\cal C}}\tilde{{\cal C}}_R\right|_0^{R_{max}} + C {\cal E},
\end{equation}
and it only follows that the constraints are zero if we control the
boundary term at $R = R_{max}$. For this reason, we impose the
momentum constraint, ${\cal C}_R = 0$, at $R = R_{max}$. This
condition results in an evolution equation for $B'$ at the outer
boundary. We combine this condition with the Sommerfeld-like
conditions at $R = R_{max}$,
\begin{equation}
\dot{A} + A' = 0, \qquad
\dot{C} + C' = 0, \qquad
\pi_\gamma + d_\gamma = 0.
\end{equation}

\subsubsection{Numerical implementation}

Next, we discuss the numerical implementation of the above constrained
evolution system. In order to apply the discretization techniques
discussed in Sect. \ref{Sect:AnNumTools} we first recast the evolution
equations into first order symmetric hyperbolic form by introducing
the new variables $\pi_A = \dot{A}$, $\pi_B = \dot{B}$, $\pi_C =
\dot{C}$ and $d_A = A' + 2G'$, $d_B = B'$, $d_C = C' + G'$. The
resulting first order system is then discretized by the method of
lines. Let us first discuss the spatial discretization which requires
special care at $R=0$ because of the coefficients proportional to
$1/R$ that appear in the evolution equations. To this end, consider
the following family of toy models
\begin{eqnarray}
\dot{\pi} &=& R^{1-n}\partial_R (R^{n-1} d),
\label{Eq:Toy1}\\
\dot{d} &=& \partial_R\pi,
\label{Eq:Toy2}
\end{eqnarray}
where $R > 0$ is the radial coordinate, and $n=1,2,3,..$. We impose
the regularity condition $d=0$ at $R=0$, which, for sufficiently smooth
fields, implies that $\dot{\pi} = n\partial_R d$ at $R=0$, and assume
that the fields vanish for $R$ sufficiently large. The toy model
(\ref{Eq:Toy1}--\ref{Eq:Toy2}) corresponds to the $n$-dimensional wave
equation for spherically symmetric solutions. The principal part of
our evolution system has precisely this form near $R=0$, where $n$
is given by $\lambda+1$, $\lambda+3$, $2$, $1$ for the evolution
equations for $\pi_A$, $\pi_B$, $\pi_C$ and $\pi_\gamma$,
respectively. The system (\ref{Eq:Toy1}--\ref{Eq:Toy2}) admits the
conserved energy
\begin{equation}
E = \frac{1}{2} \int_0^\infty R^{n-1} \left( \pi^2 + d^2 \right) dR.
\end{equation}
A second order accurate and stable numerical discretization of the
system (\ref{Eq:Toy1}--\ref{Eq:Toy2}) can be obtained as follows: We
assume a uniform grid $R_j = j\Delta R$, $j=0,1,2...$, approximate the
fields $\pi$ and $d$ by grid functions $\pi_j = \pi(R=R_j)$, $d_j =
d(R=R_j)$, and consider the semi-discrete system
\begin{eqnarray}
   \dot{\pi}_j = R_j^{1-n} D_0( R^{n-1} d)_j\quad\hbox{for $j > 0$ and } 
&& \dot{\pi}_0 = \frac{n}{\Delta R}\, d_1\; ,\\
   \dot{d}_j = D_0 \pi_j\quad\hbox{for $j > 0$ and }
&& \dot{d}_0 = 0\; ,
\end{eqnarray}
where for a grid function $u_j$, $(D_0 u)_j = (u_{j+1} -
u_{j-1})/(2\Delta R)$ is the second order accurate centered
differencing operator. It is not difficult to check that this scheme
preserves the discrete energy
\begin{equation}
E_{discrete} = \frac{\Delta R}{2} \sum_{j=1}^\infty R_j^{n-1} \left( \pi_j^2 + d_j^2 \right) 
 + \frac{\Delta R}{4n} R_1^{n-1} \pi_0^2
\end{equation}
which proves the numerical stability of the semi-discrete system.
Finally, we use a third order Runge-Kutta algorithm in order to
perform the integration in time. By a theorem of Levermore~\cite{tadmor}, 
this guarantees the numerical stability of the fully
discrete system for small enough Courant factor.

We apply these techniques for the discretization of our coupled
system. The outer boundary conditions are implemented by a projection
method. Of course, the resulting system is much more complicated than
the simple toy model problem presented above, and we have no {\it a priori}
proof of numerical stability. Nevertheless, we find the above analysis
useful as a guide for constructing the
discretization. Our resulting code is tested by running several
convergence tests, and its accuracy is tested by monitoring the
constraint variables ${\cal C}$ and ${\cal C}_R$ and the quantity
$\left. \dot{A} - (\lambda+1)\dot{B} \right|_{R=0} = 0$.

\subsubsection{Results}

Here we discuss the results for the numerical evolution of the initial
data defined by
Eqs.~(\ref{Eq:BubbleMetric}--\ref{Eq:IDU}).
We start by reviewing the evolution of the initially expanding bubbles
and the initially collapsing negative mass bubbles~\cite{SL-Rapid} and
then focus on the initially collapsing positive mass bubbles.

\paragraph{Brill-Horowitz initially expanding case}

\begin{figure}[htb]
\centerline{
\includegraphics[width=8cm]{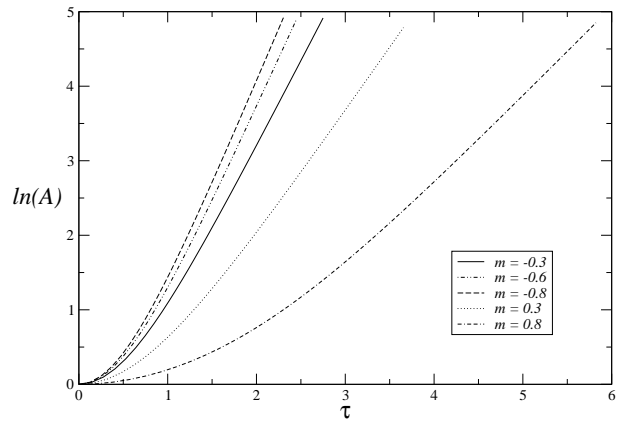}}
\caption{Bubble area vs. proper time at the bubble. In this and the
following plots, we set $r_+ = 1$. The figure shows four illustrative
examples of bubbles whose initial acceleration is positive. As it is
evident, the expansion of the bubble continues and the difference is
the rate of the exponential expansion. The relative error in these
curves, estimated from the appropriate Richardson extrapolated solution in 
the limit $\Delta \rightarrow 0$, is well below 0.001\%.}
\label{fig:area}
\end{figure}

The Brill-Horowitz initial data ($n=2$) in the case of vanishing
electromagnetic field is evolved. The bubble area $A$ as a function of
the proper time $\tau$ at the bubble is shown in Fig. \ref{fig:area}
for different values of the mass parameter $m$. As expected, the lower
the mass of the initial configuration, the faster the expansion.
Empirically, and for the parameter ranges used in our runs, we found
that at late times the expansion rate obeys
\begin{equation}
\frac{\dot{A}}{A} \approx \frac{2-\bar{m}}{r_+(\tau=0)}\, ,
\label{simplerelation}
\end{equation}
where a dot denotes the derivative with respect to proper time
$\tau$. In particular this approximation is valid for the bubble
solution exhibited by Witten~\cite{witten} which describes the time
evolution in the case $\bar{m}=0$.\\

\paragraph{Collapsing negative mass case}

\begin{figure}[htb]
\centerline{
\includegraphics[width=8cm]{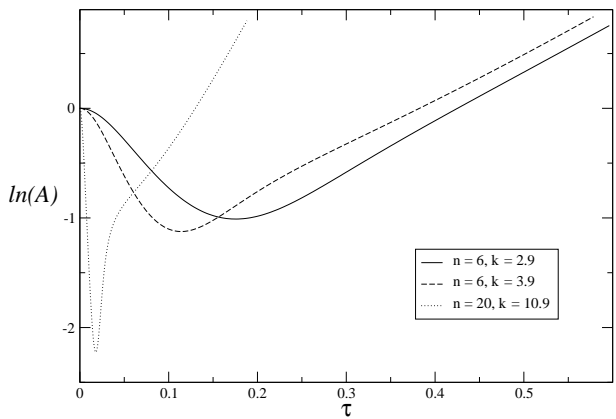}}
\caption{Bubble area vs. proper time at the bubble. The figure shows
three illustrative examples of bubble with negative mass ($m = -0.1$
each) whose initial acceleration is negative. As it is evident, the
collapse of the bubble is halted and the trend is completely
reversed. The error in these curves is estimated to be well below
0.001\%. }
\label{fig:notnakedIII}
\end{figure}

We here restrict to cases with negative masses that start out
collapsing. Interestingly enough we find that even when starting with
large initial negative accelerations, which in turn make the bubble
shrink in size to very tiny values, it bounces back without ever
collapsing into a naked singularity. As an example, Fig.
\ref{fig:notnakedIII} shows the bubble's area versus time for
different values of $n$ and $k$. The initially collapsing bubbles
decrease in size in a noticeable way but this trend is halted and the
bubbles bounce back and expand. Although we have not found a simple
law as that in Eq.~(\ref{simplerelation}), clearly the bubbles expand
exponentially fast. Therefore, it seems not to be possible to
``destroy'' the bubble and create a naked singularity. This situation
is somewhat similar to the scenarios where one tries to ``destroy'' an
extremal Reissner-Nordstr\"om black hole by attempting to drop into it
a test particle with high charge to mass ratio. There, the
electrostatic repulsion prevents the particle from entering the 
hole~\cite{wald}.

\paragraph{Brill-Horowitz initially collapsing case}

Next, we analyze the Brill-Horowitz initial data for the case in which
the bubble is initially collapsing (notice that for $n=2$ this implies
that the ADM mass is positive). While our numerical simulations reveal
that in the absence of the gauge field such a bubble continues to
collapse, we also show that when the gauge field is strong enough, the
bubble shrinks at a rate which decreases with time and then bounces
back.

\begin{figure}[htb]
\centerline{
\includegraphics[width=8cm]{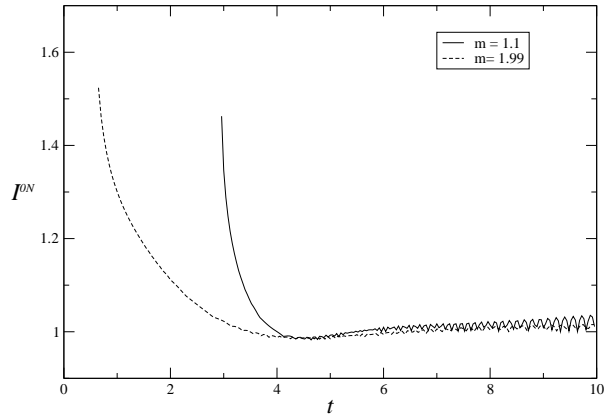}}
\caption{Rescaled Kretschmann invariant $I^{0N} \equiv I r_{AH}^4/12$ 
vs. asymptotic time for $m=1.1$ (solid line) and $1.99$ (dashed line). The first non-zero
values of the lines mark the formation of the apparent horizon. After some
transient period, both lines approach the value of $1$ suggesting a black
string has formed.}
\label{fig:blackstring}
\end{figure}

Obviously, if the collapse trend were not halted, a singularity should
form at the origin. Since the ADM mass is positive, one expects this
singularity to be hidden behind an event horizon, and one should
obtain a black string. In fact, for the solutions which are initially
collapsing and which have vanishing gauge field, we observe the
formation of an apparent horizon. Furthermore, we compute the
curvature invariant quantity $I r_{AH}^4$ at the apparent horizon (as
discussed in~\cite{choptuikblackstring}), where $I = R_{abcd}R^{abcd}$
is the Kretschmann invariant and $r_{AH}$ the areal radius of the
horizon. For a neutral black string, this invariant is $12$.  Figure
\ref{fig:blackstring} shows how this value is attained after the
apparent horizon forms for representative vacuum cases (with $m=1.1$
and $m=1.99$) this, together with the formation of apparent horizons, provides
strong evidence for the formation of a black string. \\

\begin{figure}[htb]
\centerline{
\includegraphics[width=12cm]{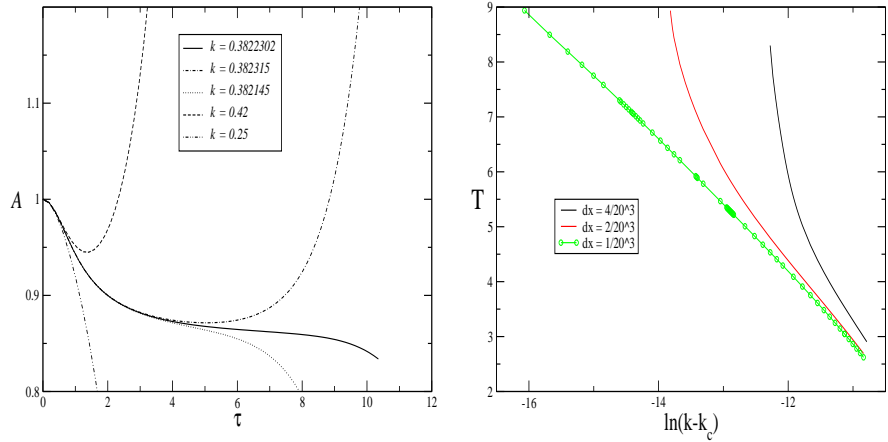}}
\caption{{\bf Left Panel.} Area values vs. proper time at the bubble for different
values of $k$ and $m=1.1$. By tuning the value of $k$ appropriately,
the amount of time that the area remains fairly constant can be
extended for as long as desired. {\bf Right Panel} The time $T$ which is a measure of how long the solution
stays close to the static solution vs. the logarithm of the difference
between the parameter $k$ and its critical value. A linear
interpolation gives the value $\gamma \approx 1.2$.}
\label{fig:criticalonset}
\end{figure}

As mentioned, for strong enough gauge fields, the previously described
dynamics is severely affected. Figure \ref{fig:criticalonset} (left
panel) shows the bubble area vs. proper time for different values of
$k$. For large values he bubble ``bounces'' back and expands while for
small ones the bubble collapses. There is a natural transition region
separating these two possibilities. Tuning the value of $k$ one can
reveal an associated critical phenomena, the `critical solution' being
a member of the family of static solutions given by
\begin{eqnarray}
&& ds^2 = -V(r) dt^2 + \frac{V(r)}{U(r)} dr^2 + \frac{U(r)}{V(r)^2} dz^2 
 + r^2 V(r) d\Omega^2,
\label{Eq:CritSol1}\\
&& \frac{1}{2}\, F_{\mu\nu}\, dx^\mu \wedge dx^\nu = \pm\frac{1}{2}\sqrt{3r_-(r_+ - r_-)}\,\frac{dr \wedge dz}{r^2 V(r)^2}\, ,
\label{Eq:CritSol2}
\end{eqnarray}
where $V(r) = 1 - r_-/r$ and $U(r) = 1 - r_+/r$. The parameters $r_-$
and $r_+$ $(> r_-)$ are related to the period of the $z$ coordinate
and to the ADM mass via $P = 4\pi r_+(1 - r_-/r_+)^{3/2}$ and $M_{ADM}
= r_+/4$. Since the quantities $P$ and $M_{ADM}$ are conserved, the
member of the family of static solutions the dynamical solution
approaches to can be determined {\it a priori} from the initial data.

Figure \ref{fig:criticalonset} (right panel) displays the time $T$
defined as the length of asymptotic time during which the bubble's
area stays within $1\%$ of the minimum value attained when the bubbles
bounces back.  This is a measure of how long the solution stays close
to the static solution as a function of the parameter $k$.
Empirically, we find the law
\begin{equation}
T = -r_+\Gamma\log | k - k_c | + T_1,
\label{Eq:CritLaw}
\end{equation}
with a parameter $\Gamma\approx 1.2$ that does not seem to depend on
the family of initial data chosen. This universality property is
reinforced by the linear stability analysis of the critical solutions
(\ref{Eq:CritSol1},\ref{Eq:CritSol2}) performed in
Ref.~\cite{SL-Stability} where we prove that each solution has
precisely one unstable linear mode growing like $\exp(\Omega t/r_+)$
with a universal Lyapunov exponent of $\Omega \approx 0.876$. This
explains the law (\ref{Eq:CritLaw}) with $\Gamma = 1/\Omega \approx
1.142$.

%
%
%
\subsection{Black holes}
\label{Sect:BH}

As one of the applications that we have chosen to illustrate the use
of the techniques previously discussed we consider here the evolution
of single non-spinning black holes. Even when the data provided
correspond to spherically symmetric and vacuum scenarios, as we will
see, obtaining a long term stable implementation is not a trivial task.  For additional
information, and a more general treatment, we refer the reader to 
Ref.~\cite{Tiglio:2003xm}.

\subsubsection{Formulation}

We adopt the symmetric hyperbolic family of formulations introduced 
in~\cite{sarbachtiglio}. This is a first order formulation whose evolved
variables are given by $\{g_{ij},K_{ij},d_{kij},\alpha,A_i\}$ with
$g_{ij}$ the induced metric on surfaces at $t=const$, $K_{ij}$ the
extrinsic curvature, $d_{kij}$ are first derivatives of the metric, 
$d_{kij} = \partial_k g_{ij}$, $\alpha$ is the lapse, and $A_i$ are
normalized first derivatives of the lapse, $A_i=\alpha^{-1}\partial_i \alpha$.

The Einstein equations written in this formulation are subject to the
physical constraints, the Hamiltonian and momentum constraints, as well
as non-physical constraints, which arise from the variable definitions.
The non-physical constraints are
\begin{equation}
C_{A_i} = A_i-N^{-1}\partial_iN = 0 \; \; , \; \; 
C_{kij} = d_{kij}-\partial_k g_{ij} = 0 \; \; , \; \;
C_{lkij} = \partial_{[l}d_{k]ij} = 0.
\end{equation}
The constraints are added to the field equations and the spacetime
{\it constraint-functions} $\{\gamma(t), \zeta(t), \eta(t), \chi(t),
\xi(t) \}$ are introduced as multiplicative factors to the constraints.
While these quantities are sometimes introduced as parameters, we
extend them to time-dependent functions.  For simplicity in this work,
we set $\zeta = -1$.  Requiring that the evolution
system is symmetric hyperbolic imposes algebraic conditions on these
factors, and they are not all independent.  If we require that all the 
characteristic speeds are ``physical'' (i.e. either normal to the spatial
hypersurfaces or along the light cone), then we obtain two symmetric hyperbolic
families.  One family has a single free parameter, $\chi(t)$,
\begin{equation}
\mbox{Single constraint-function system}
\left\{
\begin{array}{ccc}
\gamma &=& -\frac{1}{2} \\
\zeta &= & -1 \\
\eta &=& 2 \\
\xi &=& -\frac{\chi }{2} \\
\chi &\ne& 0
\end{array}
\right. \label{monoparametric}
\end{equation}
and another symmetric system with two
varying constraint-functions $\{ \eta(t), \gamma(t) \neq -1/2 \}$:
\begin{equation}
\mbox{Two constraint-function system}
\left\{
\begin{array}{ccc}
\zeta &= & -1 \\
\chi &=& -\frac{\gamma (2-\eta)}{1+2\gamma} \\
\xi &=& -\frac{\chi}{2} + \eta -2 \\
\gamma &\ne & -\frac{1}{2} \\
\eta & &
\end{array}
\right. \label{biparametric}
\end{equation}

\subsubsection{Initial data and boundaries}

Initial data for a Schwarzschild black hole are given 
in In-going Eddington--Finkelstein coordinates. The shift $\beta^i$ will be
considered an {\it a priori} given vector field while the lapse is evolved
to correspond to the time harmonic gauge with a given source
function. This gauge source function is taken from the exact solution, such
that in the high-resolution limit $\alpha=(1+2M/r)^{-1/2}$.

Black hole excision is usually based on the assumption that an inner
boundary (IB) can be placed on the domain such that information
from this boundary does not enter the computational domain.  This
requirement places strenuous demands on cubical excision for a
Schwarzschild black hole in Kerr-Schild, Painlevee-Gullstrand or
the Martel-Poisson~\cite{martelpoisson} coordinates:  the cube must
be inside $0.37M$ in each direction.  This forces one to excise
very close to the singularity, where gradients in the solution can
become very large, requiring very high resolution near the excision
boundary to adequately resolve the solution.  This requirement
follows directly from the physical properties of the Schwarzschild
solution in these coordinates, and is independent of the particular
formulation of the Einstein equations~\cite{paperIII}.

With our current uniform Cartesian code, however, we do not have
enough resolution to adequately resolve the Schwarzschild solution
near the singularity.  Thus, we place the inner boundary inside the
event horizon, but outside the region where all characteristics are
out-going.  The difference stencils are one-sided at the inner boundary,
and no boundary conditions are explicitly applied.
Testing various locations we find that placing the inner
boundary at $1.1M$ gives reasonable results for the resolutions we
are able to use, $\triangle x = \triangle y = \triangle z =  M/5,\, M/10,\, M/20$.  We are working
to resolve this inconsistency in our code by using coordinate systems
that conform to the horizon's geometry.

We performed numerical experiments with the outer boundary at three
different locations, $5M$, $10M$ and $15M$.  Boundary conditions for
the outer boundary are applied using the orthogonal projection technique
referenced above, by ``freezing'' the incoming characteristic modes. That is,
their time derivative is set to zero through an orthogonal projection. This makes use of the fact that one
knows that the continuum, exact solution is actually stationary. While this
would not be useful in the general case, as we shall see, even in such a
simplified case the constraint manifold seems to be unstable. We are currently
working on 
extending the boundary treatment to
allow for constraint-preserving boundary conditions and studying the well
posedness of the associated initial-boundary value problem. 

Having set up consistent initial and boundary data, in a second
order accurate implementation using the techniques mentioned in
section II, we now concentrate on simulating a stationary black hole
spacetime. As we will see below, even in this simple system, one
encounters difficulties to evolve the system for long times. In particular, as has been illustrated
in several occasions, the length of time during which a reliable
numerical solution is obtained varies considerably depending on the
values of the free parameters in the formulation. These parameters 
play no role  at the constraint surface; however, off this constraint
surface, these parameters have a sensible impact. Hence, at the
numerical level --where generic data is only approximately at this
surface--, it is necessary to adopt preferred values of these
parameters. These, in turn, will depend not only on the physical
situation under study but also in the details of the particular
implementation (order of convergence, etc).  As we argued in Section~II, 
the constraint minimization method provides a practical way to
adopt these parameters. We next illustrate this  in numerical
simulations of Schwarzschild spacetime.

\subsubsection{Testing constraint minimization}

We concentrate here on black hole simulations performed using the symmetric
hyperbolic formulation with two constraint functions. The single function family
and its disadvantages for constraint minimization are discussed in
Ref.~\cite{Tiglio:2003xm}.

\subsubsection{Black hole numerical results}

\begin{figure}[ht]
\begin{center}
\includegraphics*[height=6cm]{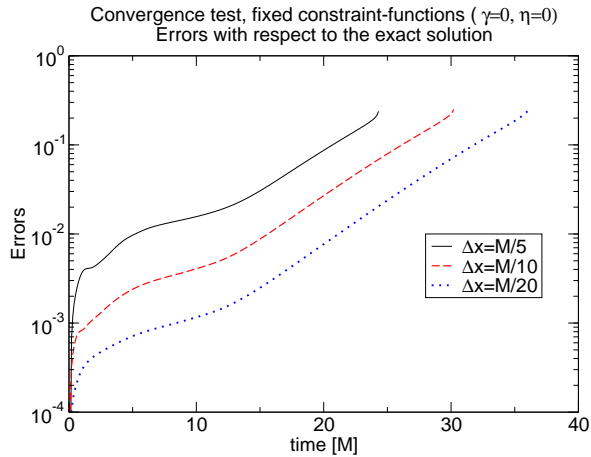}
\caption{Two-constraint-function family, with fixed values $\gamma=0=
\eta$, inner and outer boundaries at $1.1M$ and $5M$, respectively.}
\label{convergence1}
\end{center}
\end{figure}

As a first attempt to numerically integrate the Einstein equations,
one could simply fix the parameters $\eta$ and $\gamma$ to constant
values.  Lacking knowledge of preferred values for these parameters
we might simply set $\eta=0$ and $\gamma=0$.  Evolutions of the
Schwarzschild spacetime for these parameter choices, however, show that
the solution is quickly corrupted, and the solution diverges.  
Figure~\ref{convergence1} 
shows the error in the numerical solution with respect
to the exact solution for three resolutions.  While the code converges,
the error at a single resolution grows without bound as a function of time.

We now apply the constraint minimization technique to evolutions of a
Schwarzschild black hole.  The constraint functions $\eta(t)$ and
$\gamma(t)$ will  now vary in time, and both will be used to 
control the constraint growth. 
With two functions we can attempt to minimize changes
in the functions themselves. This is advantageous because smoothly
varying functions seem to yield better numerical results.
Thus, $\eta(t)$ and $\gamma(t)$ are chosen at time step $n+1$ to
minimize the quantity
\begin{equation}
\triangle := \left[ \eta(n+1) - \eta(n) \right ]^2
     + \left[ \gamma (n+1) - \gamma(n) \right]^2 \label{triangle}
\end{equation}
$\dot{\cal N}$ is nonlinear in $\gamma$ but linear in $\eta$, allowing
one to solve for $\eta$ such that $\dot{\cal N} = -a{\cal N}$,
\begin{equation}
\eta  = \frac{ - ( a{\cal N} + {\cal I}^{hom} +
  {\cal I}^{\gamma}\gamma)(1+2\gamma )  + 2\gamma
{\cal I}^{\chi}}{{\cal I}^{\eta} (1+2\gamma ) +
 \gamma {\cal I}^{\chi}}  \label{eta_tent}
\end{equation}
where, as in Section III, $a$ is given by Eq.~(\ref{a}).
$\gamma$ is chosen from some arbitrary, large interval. The
corresponding $\eta$ given by Eq.~(\ref{eta_tent}) is
computed, and the pair $(\eta, \gamma )$ that minimizes $\triangle $
defined in Eq.~(\ref{triangle}) is chosen.  $\gamma$ and $\eta$ may
be freely chosen, except that $\gamma \neq -1/2$, giving two ``branches'':
$\gamma$ always larger than -1/2, and $\gamma$ always smaller than -1/2.
We have only explored the $\gamma < -1/2$ branch using the seed
values $\eta = 0$, $\gamma =-1$.
In order to keep the variation of the parameters between two consecutive timesteps
reasonably small, we have needed to set the tolerance value for the
constraints energy roughly
one order of magnitude larger than the initial discretization error, and
$n_a$ to either $10^2$ or $10^3$. This means that the constraints' energy,
though in a longer timescale, will still grow. 

\begin{figure}[ht]
\begin{center}
\includegraphics*[height=6cm]{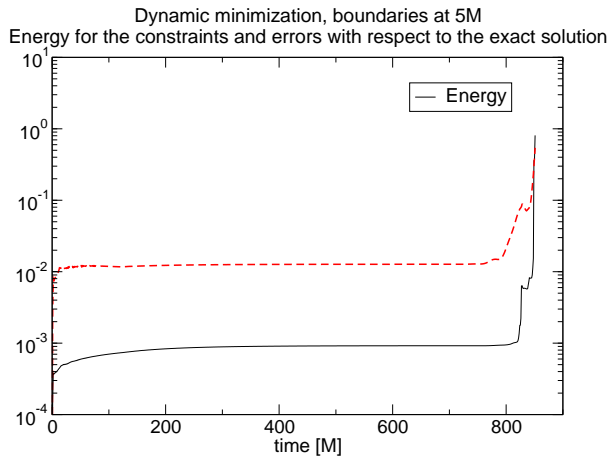}
\caption{This figure shows the constraint energy and the error
with respect to the exact solution.  Dynamic minimization is done 
with boundaries at $5M$, $\triangle x =M/5$,
$T=10^{-3}$, and $n_a=10^3$.  }
\label{bi_dyn_bound5m}
\end{center}
\end{figure}

\begin{figure}[ht]
\begin{center}
\includegraphics*[height=6cm]{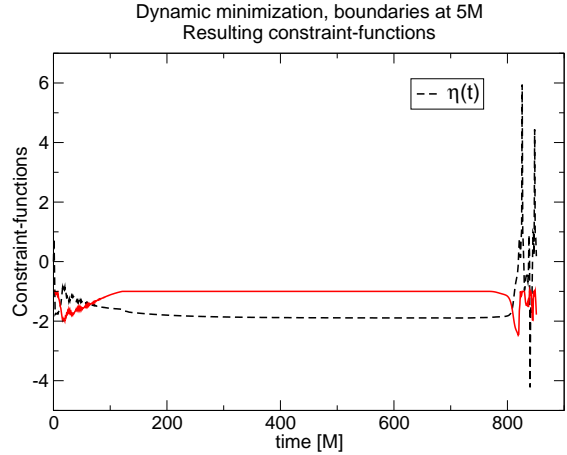}
\caption{This figure shows the constraint functions for the run
described in Figure~\ref{bi_dyn_bound5m}.
}
\label{bi_dyn_bound5mcf}
\end{center}
\end{figure}

\begin{figure}[ht]
\begin{center}
\includegraphics*[height=6cm]{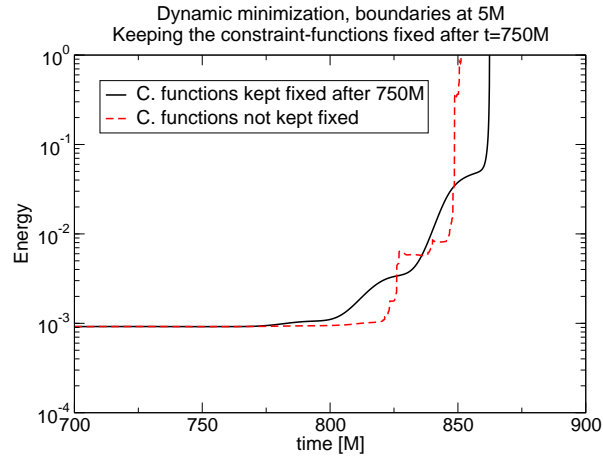}
\caption{Same as previous Figure, but keeping the constraint-functions constant
  after $750M$. The figure compares the resulting energy
  for the constraints with that of the previous figure (shown at late
  times only,  since because of the setup the runs are identical up to $t=750M$).}
\label{bi_dyn_bound5ma}
\end{center}
\end{figure}

The outer boundary is first placed at $5M$.
Figure~\ref{bi_dyn_bound5m} shows the energy of the constraints and
the error with respect to the exact solution.  The corresponding
constraint functions are shown in Figure~\ref{bi_dyn_bound5mcf}.
The large variation in the functions near the end of the run appears
to be a consequence of other growing errors.   In Figure~\ref{bi_dyn_bound5ma}
the minimization is stopped at $750M$, and the functions are fixed
to $\eta=-1.88$, $\gamma=-1.00$ for the remainder of the run.  The
solution diverges at approximately the same time.

\begin{figure}[ht]
\begin{center}
\includegraphics*[height=6cm]{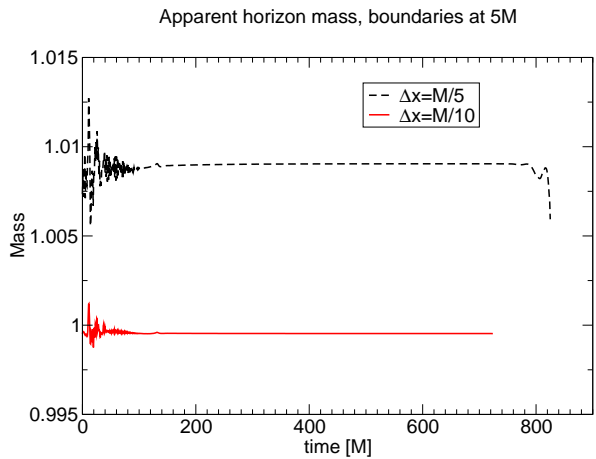}
\caption{This figure shows the black hole mass calculated from the
apparent horizon with dynamic constraint-function values.
The higher resolution simulation ran out of
computing time.  The apparent horizons were found using Thornburg's
apparent horizon finder~\cite{thornburg}.
}
\label{ah}
\end{center}
\end{figure}

Another measure of the error in the solution is the mass of the apparent
horizon, as shown in  Figure~\ref{ah}.  After some time, 
the mass approximately settles down to a value
that is around $1.009M$, which corresponds to an error of the order of
one part in one thousand.  For the higher resolution, the apparent
horizon mass at late times becomes indistinguishable from $1M$, given
the expected level of discretization errors.

\begin{figure}[ht]
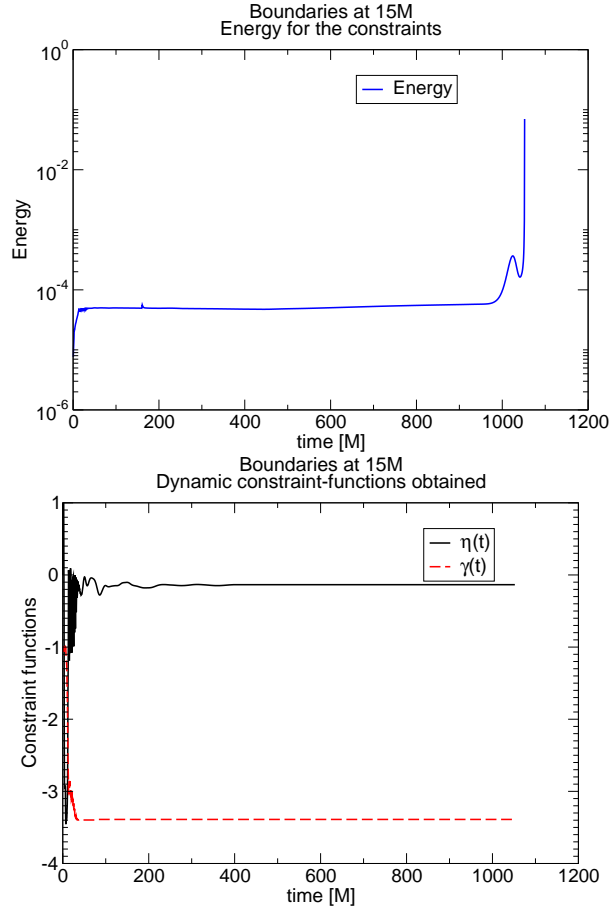

\begin{center}\includegraphics*[height=6cm]{energy_b15m.eps}
\includegraphics*[height=6cm]{pars_b15m.eps}
\caption{Dynamic
minimization done with boundaries at $15M$, $\triangle x=M/5$, $T=10^{-5}$,
and $n_a=10^2$. The constraint-functions are constant for $t \ge
450M$, where they are $\eta = -0.135, \gamma =  -3.389$. Thus, the
constraint functions do not respond when the code is about
to crash.}
\label{bi_dyn_bound15m}
\end{center}
\end{figure}

The outer boundary is now placed at $15M$.  Figure~\ref{bi_dyn_bound15m}
shows results for data equivalent to those discussed for
Figure~\ref{bi_dyn_bound5m}. 
The initial discretization value
for the energy is $7.6459 \times 10^{-6}$, and $T=10^{-5}$, $n_a=100$
was used.  The minimization of the constraint-functions is stopped
at $450M$, at which point the constraint-functions are approximately
constant, and equal to
\begin{equation}\eta = -1.35\times 10^{-1}
\;\;\; , \;\;\; \gamma =  -3.39. 
\label{pars15m}
\end{equation}

Figure~\ref{bi_dyn_bound15m} shows that the dependence of the lifetime
on the location of the outer boundaries is not monotonic, as for
this case the code runs for, roughly, $1000M$, while with boundaries at
$10M$ and $5M$ it ran for around $700M$, and $800M$, respectively.
A detailed analysis of such dependence would be computationally
expensive and beyond the scope of this work, and may even depend on
the details of the constraint minimization, such as the values for
$T$ and $n_a$. However, comparing Figure 5 with Figures
6--9, we see that the constraint minimization considerably 
improves the lifetime of the simulation, as expected. 

\section{Final Words}

We have chosen two problems to illustrate both the power of numerical
simulations of Einstein's equations and some of the difficulties
encountered in obtaining accurate numerical solutions. This is
especially relevant for black hole systems, where different poorly
understood issues coupled to lack of sufficient computational power
makes it much more difficult to advance at a sustained pace towards
the final goal of producing a reliable description of a binary black
hole system. However, it is clear that goal outweighs these
difficulties. As the bubble problem illustrates, a robust
implementation was not only key to responding to open questions but also
proved to be the way to observing other phenomena not previously considered.
Not only did it show that {\it a priori} possible way to violate
cosmic censorship is invalid, but it also revealed the existence of
critical phenomena, which, in turn, can be used to shed further light
in the stability of black string systems~\cite{SL-Stability}.

Fortunately, a substantial body of work in recent years has begun
to address a number of these questions.
A better understanding of the initial
boundary-value problem in general relativity, 
advances in the definition of initial data
and gauge choices coupled to several modern numerical techniques
 are having a direct impact in current
numerical efforts. It seems reasonable to speculate that if this
trend continues, the ultimate goal will be within reach in a 
not-too-distant future.


\acknowledgement 
This research was supported in part by the NSF under Grants No:
PHY0244335, PHY0326311, INT0204937 to Louisiana State University, the
Research Corporation, the Horace Hearne Jr. Institute for Theoretical
Physics, NSF Grant No. PHY-0099568 to Caltech, and NSF Grants No. PHY0354631
and PHY0312072 to Cornell University. This research used the
resources of the Center for Computation and Technology at Louisiana State
University, which is supported  by funding from the Louisiana legislature's
Information Technology Initiative. We thank Gioel
Calabrese, Rob Myers, Jorge Pullin and Oscar Reula for several
discussions related to the applications presented in this work.


\printindex

\end{document}